\documentclass[12pt]{iopart}

\usepackage{graphicx}
\begin{document}

\title{Poincar\'{e} Bessel  Beams: Structure and Propagation}
\author{Brianna M. Holmes and Enrique J. Galvez}

\address{Department of Physics and Astronomy, Colgate University, Hamilton, New York 13346, U.S.A.}
\ead{egalvez@colgate.edu}
\vspace{10pt}
\begin{indented}
\item 
\today
\end{indented}

\begin{abstract}
Non-separable superpositions of polarization and spatial modes in Gaussian beams have shown rich patterns of spatially-variable polarization encoded in the transverse mode of the light, with the potential for many applications. In this work we investigate encoding similar types of superpositions in Bessel beams. We find that analogously to what is observed in Poincar\'e-Gaussian beams, Poincar\'e-Bessel beams can be produced by Bessel-mode superpositions. They show polarization patterns with disclinations in the orientation of the polarization ellipse and mappings of the Poincar\'e sphere onto the transverse mode. We report on the observation of lemons, stars and monstar disclinations in the multi-ringed profile of Bessel modes. By manipulating the angular spectrum of the component Bessel beams we are able to impart linear rotation of the patterns along the non-diffracting range due to a linearly increasing propagation phase difference.
 \end{abstract}

%
\vspace{2pc}
\noindent{\it Keywords}: Bessel beams, Poincar\'{e} modes, polarization, disclinations, spatially-variable polarization, lemon, star, monstar

{\em To appear in J. Opt. (2019)}

%
%

\section{Introduction}

In recent years, there has been considerable interest in optical beams with spatially variable polarization. Besides their intrinsic interest \cite{DennisPO09,KhajaviJO16}, they are being used in a number of applications such as optical manipulation, imaging, and communications \cite{RubinszteinJO16}.
The use of spatially-variable polarization patterns has much potential for applications. Most applications involve the most studied spatially-variable beams, known as vector beams. They have been used in applications such as optical manipulation \cite{HuangOL12,ZhongAPL14,GuAO15}, micromachining \cite{MeierAPA07,MatsusakaOL18}, and communication  \cite{MilioneOL15,VallonePRL14,ZhangIEEE}.  
There is increasing interest on polarization structures of Bessel beams due to their non-diffracting character \cite{DurninPRL87,McLoinCP05}, which have found wide-ranging applications such as imaging \cite{PlanchonNM11}, microscopy \cite{FahrbachNP10}, quantum effects \cite{McLarenNC14}, and propagation through turbulent media \cite{ChengOE16}. Thus, a number of recent studies have reported on the generation of vector-Bessel beams using various methods \cite{NivOL04,FloresPerezOL06,ItoJOSAA10,DudleyOL13,SchimpfOE13,VyasOL14,WuPRA14,MilioneJO15,FuSR16}, and applications in manipulation have already been reported \cite{NiePLA15}. Poincar\'{e} modes are the general case of vector modes, because they feature spatially variable polarization of all the states in the Poincar\'e sphere \cite{BeckleyOE10,CardanoAO12,GalAO12}. Their specific application is present in problems such as chiral separation, where polarization helicity gradient imparts forces on chiral molecules \cite{Canaguier2013,Bradshaw2014,Cameron2014,KravetsPRL19}; in the manipulation of azobenzenes in liquid crystals \cite{SlussarenkoOE11}; and in the ablation of materials with structured polarization \cite{NivasSR17}.

Studies of Poincar\'{e}-Bessel modes have been limited \cite{ShvedovOE15}. 
Taking advantage of our method of using intra-beam superposition with dual spatial light modulators \cite{KhajaviOE15}, we study the recreation of the polarization disclinations done in Poincar\'{e}-Gaussian beams. We also investigated extending the variability of the polarization along the longitudinal direction \cite{SchulzePRA15,MorenoOL15,DavisOL16,LiOL16,LiOE17}, which in the case of Poincar\'{e} modes can manifest as a rotation in the pattern with propagation.  

The structure of the article is to present the mathematical description behind the experiments in Sec.~2. It is followed by a presentation of the apparatus in Sec.~3. Results and comparisons with theoretical modelings are presented in Sec.~4. In Sec.~5 we present summary and conclusions.

\section{Theoretical Considerations}
We present two types of experiments. In a first kind we study the disclinations in the polarization, consisting of lemons, stars and monstars \cite{BerryHannayJPA77}. In the second kind we investigate propagation effects.
\subsection{Poincar\'e Bessel Modes}

To prepare a mode with a desired polarization disclination, we combined coherent Bessel-modes superpositions in orthogonal states of circular polarization:
\begin{equation}
        U=e^{ik_zz}\left[e^{i\ell_1\phi}J_{|\ell_1|}(k_tr)e^{i\delta}\hat{e}_L + 
        \left(\cos\beta e^{i\ell_2\phi}
        + \sin\beta e^{-i\ell_2\phi}e^{i\gamma}\right)J_{|\ell_2|}(k_tr)\hat{e}_R\right]
\label{eq:disc}
\end{equation}
where $\beta$, $\gamma$, and $\delta$ are controllable phases; $k_t$ and $k_z$ are the transverse and axial wave vectors, respectively; and $J_{|\ell_1|}$ and $J_{|\ell_2|}$ are Bessel functions; $(r,\phi,z)$ are the cylindrical coordinates; $\ell_1$ and $\ell_2$ are integers specifying the topological charges; and $\hat{e}_R$ and $\hat{e}_L$ denote the right and left circularly polarized states, respectively. 
When $\beta=0$, the patterns produced angularly symmetric disclinations: either a lemon or star pattern depending on the sign of the topological charges. Non-zero values of $\beta$ allowed us to generate asymmetric monstar disclination patterns.

\subsection{Disclination Patterns}
The patterns produced by nonseparable superpositions of spatial modes and polarization contain disclinations, or dislocations in the rotational order. Disclinations are encoded in the orientation of the polarization ellipses \cite{FreundOC02}. Disclinations can be characterized by their index, which tells the number of turns the polarization ellipses makes about the central singularity. If $\theta$ is the orientation of the polarization ellipse, then the index is defined as \cite{KhajaviJO16}:
\begin{equation}
    I_C = \frac{1}{2\pi}\oint d\theta 
    \label{eq:ict}
\end{equation}
where the integral is over a closed path around the center of the pattern that contains the singularity, which for the case of ellipse fields is known as the C-point \cite{NyePRSLA83b,FreundOC02}; a point of circular polarization, and hence of undefined orientation. Implicit in Eq.~\ref{eq:ict} is the dependence of $\theta$ on $\phi$. It's useful to also define disclinations of the ellipse orientations in terms of their angle relative to the radial direction:
\begin{equation}
    \theta_r=\theta-\phi,
    \label{eq:thetar}
\end{equation}
which we can use to redefine the disclination index:
\begin{equation}
    I_C=1+\frac{1}{2\pi}\oint d\theta_r. 
    \label{eq:icr}
\end{equation}
Disclination patterns of the kind presented here consist of sets of curved lines that  divide a circular region into pie sectors. When the pattern is angularly symmetric these pie sectors have the same size. The sectors are delimited by radial lines. We can define radial lines by their orientation: $\theta_r=n\pi+\phi_0$ (where $n$ is an integer and $\phi_0$ is a constant). The number of radial lines can be obtained by:
\begin{equation}
    N=\frac{1}{\pi}\left|\oint d\theta_r\right| .
    \label{eq:Ndef}
\end{equation}
By relating equations (\ref{eq:icr}) and (\ref{eq:Ndef}) we can express the number of radial lines in terms of the disclination index:
\begin{equation}
    N = |2(I_C-1)| .
    \label{eq:6}
\end{equation}
Because the ellipse orientation $\theta$ is half of the relative phase between right and left circularly polarized components \cite{GalAO12}, the orientation can be described by
\begin{equation}
    \theta = \frac{\Delta \ell \phi - \delta}{2},
    \label{eq:7}
\end{equation}
where  $\delta$ is the phase between the two polarization components and $\Delta\ell=\ell_1-\ell_2$. Replacing Eq.~\ref{eq:7} into the definition, Eq.~\ref{eq:ict}, leads to
\begin{equation}
    I_C=\frac{\Delta \ell}{2}
    \label{eq:icdel}
\end{equation}
and from Eq.~\ref{eq:6},
\begin{equation}
    N = |\Delta\ell-2| .
    \label{eq:nic}
\end{equation}
The above relations describe two types of disclinations: lemons, for which $I_C>0$ and stars, for which $I_C<0$ \cite{NyePRSLA83b}. The relations are broken by asymmetric disclinations known as monstars \cite{BerryHannayJPA77,FreundOC02,DennisOC02,GalPRA14}. These asymmetric disclinations can have characteristics of both lemons and stars \cite{KhajaviJO16,GalJOSAA17}. 

Lemons, and their high-order versions sometimes called hyperlemons contain sectors that can have 3 types of lines \cite{Firby}: hyperbolic, where lines avoid the (central) singularity; elliptic, where lines start and end at the singularity; and parabolic, where only one end of the line connects to the singularity. Stars have sectors with only hyperbolic patterns, and monstars have sectors with parabolic and either elliptic or hyperbolic sectors \cite{GalJOSAA17}.  An elliptic sector has an index of $+1/2$ that contributes to the total index, because relative to the radial direction the tangent to the lines makes half a turn clockwise when going  clockwise from radial line to radial line. A hyperbolic sector contributes with an index of $-1/2$ via counter-clockwise tangent rotations for the same interval. Parabolic sectors do not contribute to the index because the tangent to disclination lines relative to the radial direction does not rotate in between radial lines. Because  of the rotational symmetry, a pattern with only parabolic lines has an index of $+1$. Bendixon first realized that the index of a pattern is given by \cite{BendixonAM1901}
\begin{equation}
I_C=1+\frac{e-h}{2},
\label{eq:icbend}
\end{equation}
where $e$ and $p$ are the number of elliptic and hyperbolic sectors, respectively.
Patterns with all of these types of lines and indices are shown in detail below.

\subsection{Pattern Rotation }
When the pattern is symmetric (Eq.~\ref{eq:disc} with $\beta=0$),
the disclination pattern's orientation is given by
\begin{equation}
    \theta_r = \frac{\Delta \ell \phi - \delta}{2} - \phi.
    \label{eq:thetrr}
\end{equation}
This orientation can be followed through the angular position of a given radial line, for which $\theta_r=n\pi$, with $n$ an integer. Thus, solving for the angular position of a radial line $\phi_r$ yields
\begin{equation}
    \phi_r = \frac{n\pi+\frac{\delta}{2}}{\frac{\Delta \ell}{2}-1}.
    \label{eq:phir}
\end{equation}
If we vary $\delta$, the pattern will rotate at a rate
\begin{equation}
    \frac{d\phi_r}{d\delta} = \frac{1}{\Delta \ell -2},
    \label{eq:rot1}
\end{equation}
or alternatively
\begin{equation}
 \frac{d\phi_r}{d\delta} =\frac{1}{2(I_C-1)}.
 \label{eq:rateic}
 \end{equation}
 The phase $\delta$ makes the ellipses at individual points in the mode to rotate, but because of the symmetry of the pattern,  the pattern  rotates. However, the beam itself does not rotate. This will not occur for non-symmetric patterns where $\beta\ne0,\pi/2$.
 
 \subsection{Rotation due to Propagation}
Bessel modes are produced by manipulating the angular spectrum of a plane wave, or by tilting the propagation vector by the angle
\begin{equation}
\alpha=\frac{k_t}{k_z}.
\end{equation}
In our assembly of the Poincar\'{e} mode in an azimuthally symmetric pattern we selected a particular value of $\alpha$. However, since the two Bessel modes are programmed separately, we can encode distinct values of $\alpha$ on each mode.
The resulting (symmetric) mode is given by
\begin{equation}
        U= e^{i(k_{z_1}z+\ell_1\phi)}J_{|\ell_1|}(k_{t_1}r)e^{i\delta}\hat{e}_L+ e^{i(k_{z_2}z+\ell_2\phi)}J_{|\ell_2|}(k_{t_2}r)\hat{e}_R .
    \label{eq:Prop1}
\end{equation}
Thus, along the region where both Bessel beams are non-diffracting, the patterns will rotate because of the difference in the axial wave-vector of the Bessel modes.
Following the same argument as in the previous section, the orientation of the pattern as defined by the orientation of a radial line will be given by
\begin{equation}
    \phi_r = \frac{n\pi+\frac{\delta}{2} -\frac{\Delta k_z z}{2}}{\frac{\Delta \ell}{2}-1}.
    \label{eq:phirz}
\end{equation}
The orientation will rotate as a function of $z$:
\begin{equation}
    \frac{d\phi_r}{dz} = \frac{\Delta k_z}{\Delta \ell -2},
    \label{eq:rotz1}
\end{equation}
or alternatively
\begin{equation}
 \frac{d\phi_r}{dz} =\frac{\Delta k_z}{2(I_C-1)}.
 \label{eq:rotzic}
 \end{equation}
Thus, there should be a linear relationship between the angular orientation of the pattern  and $z$. 

\section{Experimental Procedure}
\begin{figure}[h]
    \centering
    \includegraphics[width=0.8\textwidth]{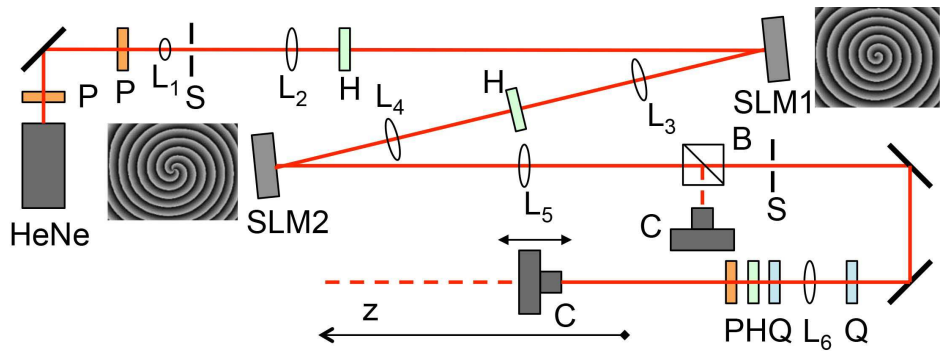}
    \caption{Schematic of the experimental apparatus. Optical components Lenses (L$_i$, $i=1$-6),  polarizers (P), half-wave plates (H), quarter-wave plates (Q), non-polarizing beam splitter (B), spatial light modulators (SLM), spatial filter (S), digital camera (C) and helium-neon laser (HeNe).}
    \label{fig:app}
\end{figure}

To create the Poincar\'e-Bessel beam, we designed the setup shown in Figure \ref{fig:app}. We sent the Gaussian beam from a helium-neon (HeNe) laser (of wavelength 632.8 nm) through a pair of polarizers to control its intensity. Two lenses (L$_1$ and L$_2$ with focal lengths $f=7.5,40$ cm, respectively) were used to expand the beam so it would fill the spatial Light Modulator (SLM). A half-wave Plate (HWP) was set to make the beam diagonally polarized so that it would have equal amplitudes of horizontally and vertically polarized light. The horizontal component was phase-modulated by the first SLM (SLM1) to produce a Bessel beam with topological charge, $\ell_1$ upon reflection. 
 After SLM1, a 4-f system of lenses (L$_3$ and $L_4$ with $f = 25$ cm) was setup to so that the beam was re-imaged onto the second SLM (SLM2). In the middle of this 4-f system a HWP set to 45$^{\circ}$ flipped the horizontal and vertical components so that SLM2 would modulate the other polarization component. A quarter-wave plate (QWP) was placed near the end of the path to convert the two polarization components into circular polarization states. The QWP in combination with the SLM's effectively encoded the light according to Eq.~\ref{eq:disc}. 

The images inserted in Fig.~\ref{fig:app} show  magnified  examples of phase programming of the Bessel modes: of order $\ell_1=1$ in SLM1  and  $\ell_2=-3$ in SLM2. Because of the angular and radial modulation, the programmed patterns are phase spirals. The number of branches in each spiral is indicative of the value of the topological charge. 
This encoding method involves superposition of orthogonal modes of the same beam \cite{KhajaviOE15}. It can also be viewed as a form of intra-beam interference, because both interfering components are part of the same beam. As such the superposition is very stable and not susceptible to vibrations or air currents.

After the SLM2 the beam traveled through another 4-f system. This second 4-f system (L$_5$ and L$_6$ with $f= 50$ cm) was used to image the  Poincar\'e-Bessel beam onto a digital camera.  We imaged the Fourier-transform by use of a beam splitter to divert part of the light to another camera. We verified that it consisted mainly of a bright ring of radius proportional to $k_t$. Additional features in the Fourier plane were spatially filtered by an aperture to improve the quality of the beam.

A polarimetry setup placed before the beam-imaging camera consisted of a QWP followed by a HWP and a Glan-Thompson polarizer set to horizontal. By changing the angles on the QWP and HWP  it was possible to image the light projected onto the 4 linear polarization eigenstates: horizontal, vertical, diagonal and antidiagonal; plus the 2 left and right circular polarization eigenstates. An image of the intensity of each of these polarization states was recorded to find the Stokes parameters and hence the state of polarization of each imaged point. 

\section{Experimental Results \& Comparisons with Theory}

In this section, we first analyze the structure of the patterns for different combinations of topological charges, and as a function of the variables of Eq.~\ref{eq:disc}: $\beta$, $\gamma$, and $\delta$. We follow with the study of the rotation of the patterns at varying distances. 

\subsection{Lemon, Star, \& Monstar Patterns}

\begin{figure}[htb]
    \centering
   \includegraphics[width=9.7cm]{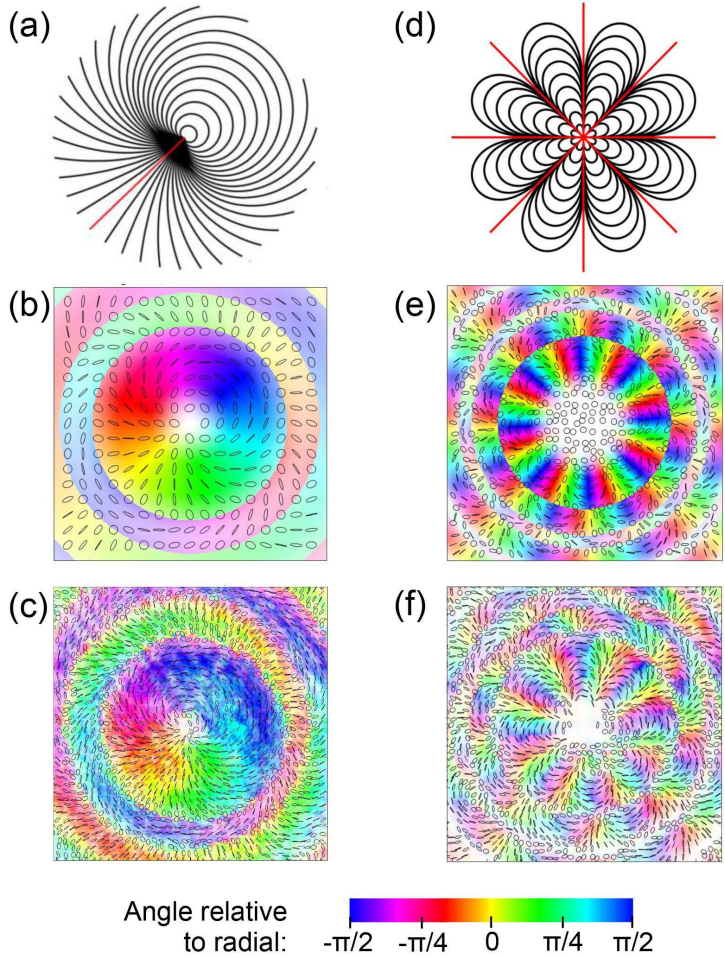}
        \caption{Modeled and measured patterns for lemons. Disclination line maps with red radial line are shown in (a,d), theoretical expectations in (b,e), and experimental results in (c,f) for two cases: (a-c) $(\ell_1,\ell_2)=(2,-1)$, $I_C=3/2$, $N=1$; and (d-f) $(\ell_1,\ell_2)=(6,-4)$, $I_C=5$, $N=8$. False color encodes the orientation of the ellipses with respect to the radial direction (yellow). Colour saturation encodes intensity.}
    \label{fig:lemon}
\end{figure}

In Fig.~\ref{fig:lemon} we show two types of lemons. In the top row (a,d) we show the  line patterns that highlight the structure of the disclination, with radial lines delimiting sectors drawn in red. The second row (b,e) and third row (c,f) correspond to the modeling and measured patterns, respectively. We display them in a way that is convenient to easily identify the patterns. We use false colour to encode the radial orientation of the polarization ellipses  (defined by Eq.~\ref{eq:thetar}), with yellow encoding the radial orientation including radial lines, and blue the azimuthal orientation. The drawn ellipses represent the local state of polarization. They are placed  at random locations to avoid any bias in the determination of the pattern. 

The disclination patterns of Fig.~\ref{fig:lemon} (a) and (d) have 1 and 8 elliptical sectors, respectively, which by use of Bendixon's relation, Eq.~\ref{eq:icbend}, we see that they correspond to $I_C=+3/2$  and $I_C=5$, respectively. The actual patterns were made by programming SLM1 and SML2 with Bessel modes with topological charges $(\ell_1,\ell_2)=(2,-1)$ for (a-c) and $(6,-4)$ for (d-f). The images were taken at the start of the non-diffracting region, at about 50 cm from the last lens (L$_6$ in Fig.~\ref{fig:app}). It can be seen that the images in (c) and (f) obtained by imaging polarimetry compare well with the  modeled ones in  (b) and (e), respectively. Notice also the abrupt change in the orientation of the ellipses at certain radii, corresponding to the $\pi$ phase flip in the Bessel modes. This is also seen by the sharp change in the colours. This is much more visible in the data in (c), but agrees qualitatively with the modeling in (b). The number of orientation flips is greater in (e,f) because both component Bessel modes, of orders 6 and 4, have phase flips at differing radii. It can be seen that the pattern of ellipse orientations flows smoothly between every other radial sector. Differences between the expectation and measurements may be due to the lack of amplitude modulation in the SLMs, although we tried to mitigate this by masking in the Fourier plane.

\begin{figure}[bht]
    \centering
       \includegraphics[width=9.7cm]{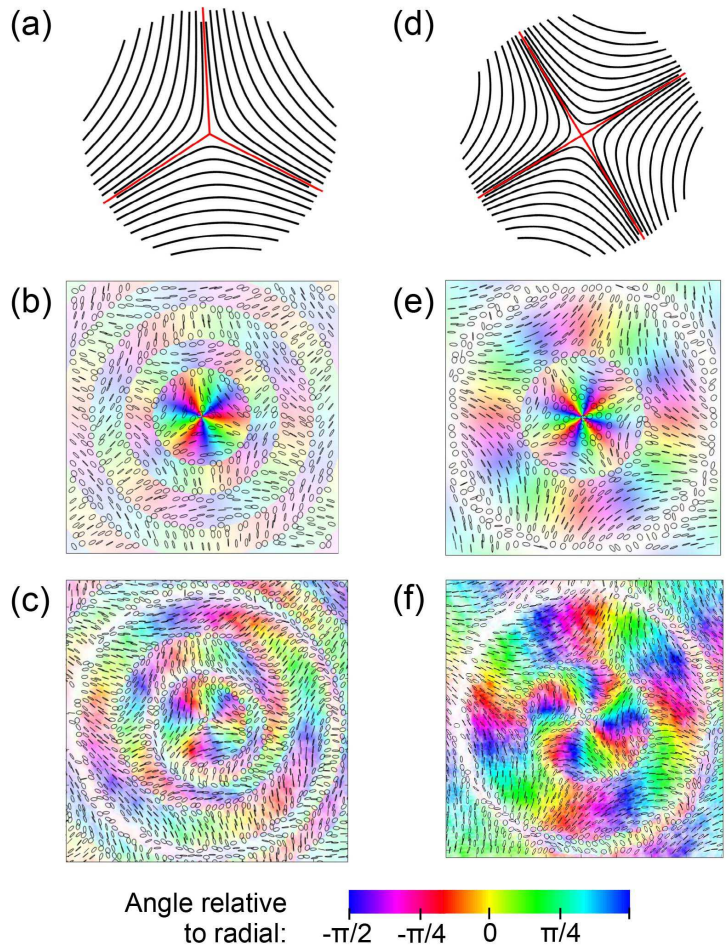}
    \hfill
        \caption{Modeled and measured patterns for stars. Disclination line maps with red radial line are shown in (a,d), theoretical expectations in (b,e), and experimental results in (c,f) for patterns: (a-c) $(\ell_1,\ell_2)=(0,1)$, $I_C=-1/2$, $N=3$; and (d-f) $(\ell_1,\ell_2)=(-2,0)$, $I_C=-1$, $N=4$. False color encodes the orientation of the ellipses with respect to the radial direction (yellow). Colour saturation encodes intensity.}
    \label{fig:star}
\end{figure}

Figure~\ref{fig:star} shows the cases of two star patterns: $I_C=-1/2$ in (a-c), and $I_C=-1$ in (d-f). The line disclinations contain only sectors with hyperbolic lines, as described previously. The comparisons between modeling and measurements are also qualitatively very good. The case of Fig.~\ref{fig:star}(f) shows prominently an interesting feature of the Bessel-mode superpositions: the zeros of the two component Bessel modes nearly coincide: the first zero of $J_2(x)$ at $x=5.1$ is near the second zero of $J_0(x)$ at $x=5.5$. This creates a broad ring of near-zero intensity, as seen as the radial region with no colour in the figure. There is also a second near-coincidence at the next zero, but it is not visible in the figure. One can envision engineering the modes to have a tubular region with a broad region of near zero intensity. 
In addition to the angular variation in the orientation of the ellipses, the ellipticity varied rapidly because it depends on the ratio of the amplitudes of the component modes. When the amplitudes are the same, the polarization is linear. At every crossing of the magnitude of the amplitudes, the handedness of the ellipse is flipped. Thus, the modes contain quite a variability in the polarization, constituting many mappings of the Poincar\'e sphere onto the transverse mode of the beam.

\begin{figure}[htb]
    \centering
      \includegraphics[width=9.7cm]{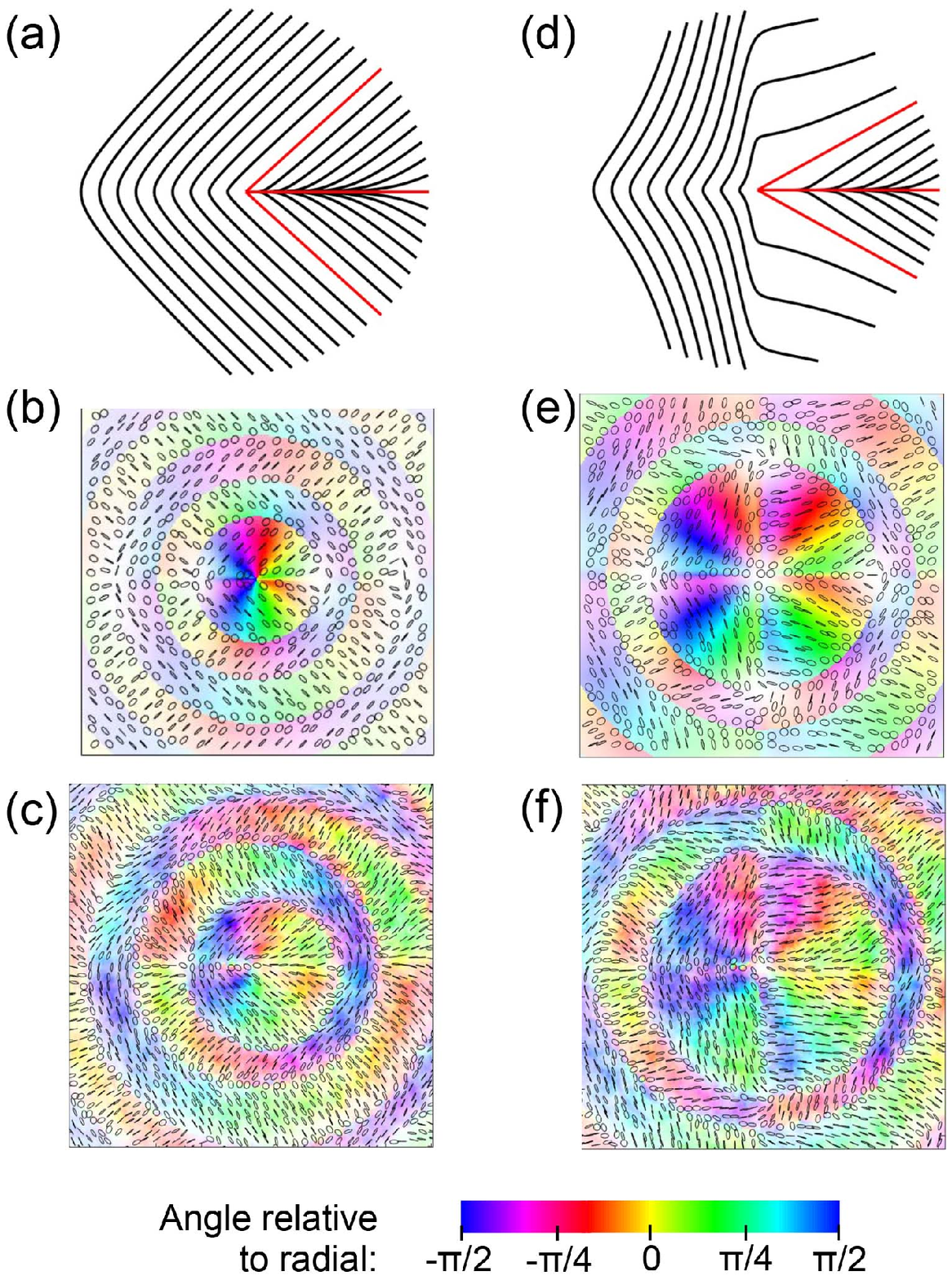}
    \hfill
        \caption{Modeled and measured patterns for monstars. Disclination line maps with red radial line are shown in (a,d), theoretical expectations in (b,e), and experimental results in (c,f) for a two monstar patterns: (a-c) $(\ell_1,\ell_2)=(1,0)$, $\beta = 40^\circ$, $\gamma = 180^\circ$; and  (d-f) $(\ell_1,\ell_2)=(2,1)$, $\beta = 40^\circ$, $\gamma = 180^\circ$. False color encodes the orientation of the ellipses with respect to the radial direction (yellow). Colour saturation encodes intensity.}
    \label{fig:monstar}
\end{figure}

In Fig.~\ref{fig:monstar} we show that asymmetric monstars can also be made in superpositions of Bessel modes. We show two cases of monstars. Both have the same index $I_C=-1/2$ but different line structure. This is typical of these recreations using modal superpositions. In the case of symmetric modes, the difference in the topological charge $\Delta\ell$ determines the index and the shape of the disclination lines, but in monstars the asymmetry makes the individual values of $\ell_1$ and $\ell_2$ also define the exact line pattern, and also the range of values of $\beta$ for which the monstar pattern is present \cite{GalJOSAA17}. As can be seen in the two cases shown, the differences are subtle but visible. The case of Fig.~\ref{fig:monstar}(a) is the classic monstar pattern envisioned originally 42 years ago by Berry and Hannay \cite{BerryHannayJPA77}. The pattern of Fig.~\ref{fig:monstar}(d) was realized only recently \cite{KhajaviJO16,GalJOSAA17} in Poincar\'{e}-Gaussian beams, but the specific pattern shown here has not been published previously. The modulations in the lines are clearly seen in the data.
The agreement between the measurements and expectation is qualitatively very good. 

\subsection{Pattern Rotation}

Earlier we described two methods for imparting rotation to symmetric patterns. In the first one, the relative phase $\delta$ was varied for a fixed position $z$. We did measurements on this, not shown here, where we imaged the beam at $z=50$ cm and then analyzed the beam's polarization for $\delta = (0,\frac{\pi}{2},\pi,\frac{3\pi}{2},2\pi)$. The rotation of the images was in excellent quantitative agreement with Eq.~\ref{eq:phir}. The ellipses at each location in the beam rotate due to the inserted phase, giving rise to a rotation of the pattern. 

\begin{figure}[htb]
    \centering
      \includegraphics[width=9.7cm]{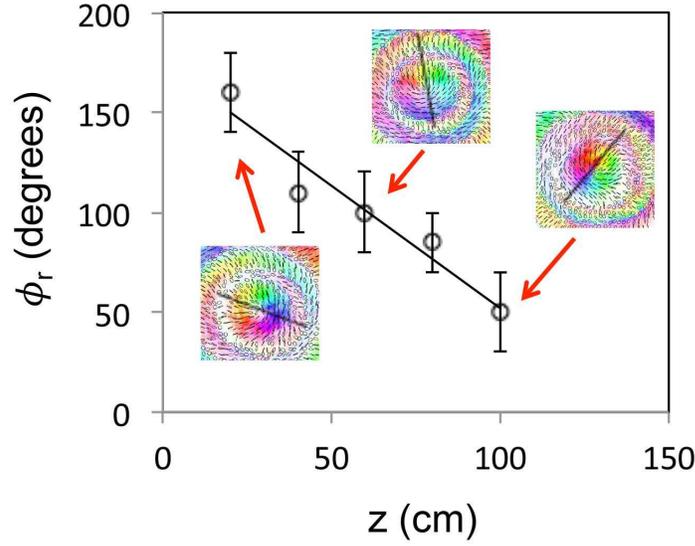}
    \caption{Graph showing the rotation of the polarization pattern as a function of $z$ for the case  $(\ell_1,\ell_2)=(0,-1)$. The inserts show the patterns for three values of $z$. The black line is used to define the orientation of the mode, which also corresponds to the radial line.}
    \label{fig:phivz}
\end{figure}

The second type of rotation involves preparing the light according to Eq.~\ref{eq:Prop1}, where the Bessel modes were programmed with different radial wave-vectors $k_t$ and consequently different axial wave-vectors $k_z$, as discussed earlier. Because the modes are encoded in states of circular polarization, the phase difference due to the different axial wave-vectors manifests in a rotation of the pattern. This type of phase has been implemented before to change the polarization of the Bessel beam as a function of $z$ \cite{SchulzePRA15,MorenoOL15}. 
We studied several cases $(\ell_1,\ell_2)$. In the cases that we investigated we chose $k_{t_1}=0.5k_{t_2}$. Correspondingly, the images obtained in the Fourier plane consisted mainly of 2 bright rings of radii differing by a factor of 2.  The agreement between the measurements and the expectations is very good. Figure~\ref{fig:phivz} shows an example of the rotation as a function of $z$ of the mode made with topological charges $(0,-1)$. It can be seen that the pattern indeed rotates as a function of the propagation distance. The fitted value of the slope is $1.23\pm0.19$. We note that we took measurements as early as we can see the whole pattern, at about 20 cm from lens L$_6$.  

\begin{figure}[htb]
    \centering
      \includegraphics[width=9.7cm]{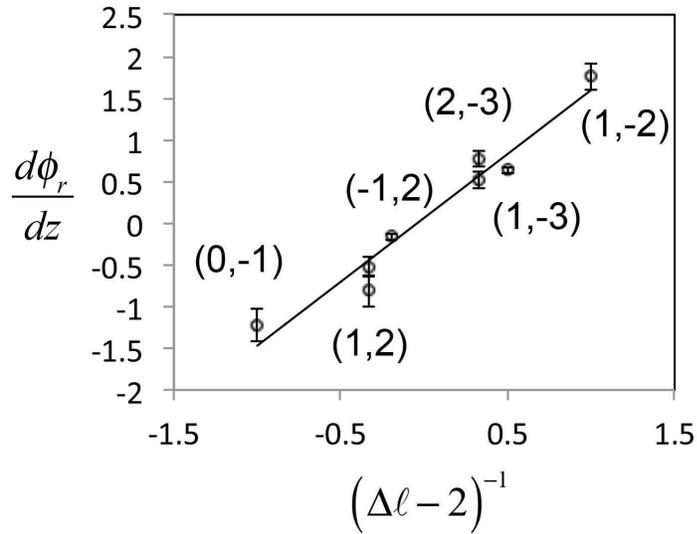}
    \caption{Graphs of the rate of change of the orientation of the patterns as a function of $\Delta\ell-2$ for distinct cases of Poincar\'{e}-Bessel modes with corresponding values of $(\ell_1,\ell_2)$ labeling the points.}
    \label{fig:slopes}
\end{figure}

We took measurements for a range of values of $(\ell_1,\ell_2)$ and measured their respective rotation. We summarize the observed rotation in Fig.~\ref{fig:slopes} by graphing the data according to Eq.~\ref{eq:rotz1}.
The data follows a linear relation that is expected when plotting the rotation as a function of $1/(\Delta\ell -2)$.

\section{Conclusion \& Discussion}
In summary, we have shown that we can prepare Bessel beams in any Poincar\'{e} mode that is desired, similar to what has been done with Poincar\'{e}-Gaussian beams. The prescription is the same: non-separable superpositions of spatial mode and polarization. We were able to generate lemon, star, and monstar patterns in agreement with the expectations. The spatially-variable patterns of polarization consisted of a central spot containing the C-point surrounded by concentric rings that were due to the structure of Bessel beams. For the lemon and star patterns, the central circle contained the expected disclination pattern around the C-point, while each adjacent ring had ellipses that were rotated by $\pi/2$ with respect to adjacent rings. Thus, every other ring followed the same orientation disclination pattern. We note that in contrast with Gaussian vortex beams, Bessel vortices are less affected from perturbations because they are always imaged in the near field. Thus one can obtain better quality Poincar\'e patterns than those obtained in the far field \cite{KhajaviOE15}. 

We also showed that by preparing Bessel-mode/polarization superpositions of Bessel modes with distinct radial wave-vectors, we can impart a rotation to the pattern of polarization. By changing the relative magnitude of the transverse wave-vectors we could vary the amount of rotation per distance of travel.
This study shows a type of control of the polarization of patterns that preserves the size of the disclination carried by the beam mode while rotating it.  This contrasts the case of Gaussian beams, where the only way to impart a longitudinal variation is through the intrinsic Gouy phase \cite{CardanoOE13}. Gaussian beam propagation also involves a variation in the size of the mode due to diffraction. The type of beams presented here  may find applications in the encoding of information in light beams, new forms of imaging and manipulation of polarization-sensitive materials. Other types of pattern manipulation may lead to interesting variations of spatially-variable polarization from the orderly, shown here, to purely random variability \cite{KhoninaOL15}.

\section{Acknowledgements}
This work was funded by National Science Foundation grant PHY-1506321.
\section{References}
\bibliography{Jo19} 
\bibliographystyle{iopart-num} 

\end{document}